\documentclass[a4paper,12pt,tightenlines,notitlepage,secnumarabic]{revtex4-1}
\usepackage{amsmath,amssymb,setspace}
\usepackage{graphicx}
\usepackage{newtxtext}
\usepackage{newtxmath}
\usepackage[normalem]{ulem}
\usepackage{xfrac,verbatim}
\usepackage[utf8x]{inputenc}
\usepackage[nodayofweek,ddmmyy]{datetime}
\usepackage{color}
\usepackage[unicode,colorlinks,bookmarks=true]{hyperref}

\newcommand{\eg}{{\it e.g.\,}}
\newcommand{\Tr}{\mathop{\rm Tr}\nolimits}
\newcommand{\bra}[1]{\langle#1|}
\newcommand{\ket}[1]{|#1\rangle}
\newcommand{\bracket}[2]{\langle#1|#2\rangle}

\begin{document}

\title{High-fidelity detection of a phase shift using non-Gaussian quantum states of light}

\author{F.Ya.Khalili}
\email{khalili@phys.msu.ru}
\affiliation{Faculty of Physics, M.V. Lomonosov Moscow State University, 119991 Moscow, Russia}
\affiliation{Russian Quantum Center, Skolkovo 143025, Russia}

\begin{abstract}

We show that by injecting a light pulse prepared in a non-Gaussian quantum state into the dark port of a two-arm interferometer, it is possible to detect a given phase shift with the fidelity which is limited only by the optical losses and the photodetection inefficiency. The value of the phase shift is inversely proportional to the amplitude of the classical carrier light injected into another (bright) port of the interferometer. It can be reduced by using an additional degenerate parametric amplifier (squeezer) in the input dark port and the matching anti-squeezer in the output dark port. 

We show that using the modern high-efficiency photon number resolving detectors, it is possible to reduce the detection error by almost one order of magnitude in comparison with the ordinary (Gaussian-state) interferometry. 

\end{abstract}


\maketitle

\section{Introduction}\label{sec:Intro}

At the most fundamental level, the sensitivity of interferometric optical phase measurements is limited by the quantum uncertainties of the probing light and therefore depends on its quantum state. In the case of coherent quantum states generated by phase-stabilized lasers the phase sensitivity corresponds to the shot noise limit: 
\begin{equation}\label{SNL}
  \Delta\phi_{\rm SNL} \sim \frac{1}{\sqrt{N}} \,,
\end{equation}
where $N$ is the mean number of photons used for the measurement. Better sensitivity, for a given $N$, can be achieved by using squeezed quantum states \cite{Caves1981}. If the  squeezing is not very strong, $e^{2r}\ll N$, where $r$ is the logarithmic squeeze factor, then the phase sensitivity can be improved by $e^r$ in comparison with the SNL:
\begin{equation}\label{dphi_sqz}
  \Delta\phi_{\rm SQZ} \sim \frac{e^{-r}}{\sqrt{N}} \,.
\end{equation}
This improvement was demonstrated in kilometers-scale interferometers of
laser gravitational-wave detectors GEO-600 and LIGO \cite{Nature_2011, Nature_2013, Grote_PRL_110_181101_2013}. In the very strong squeezing case of $e^{2r}\sim N$, the phase sensitivity of squeezed states reaches the Heisenberg limit \cite{Yurke_PRA_A_33_4033_1986, Ou_PRL_77_2352_1996, 17a1MaKhCh}:
\begin{equation}\label{HL}
  \Delta\phi_{\rm HL} \sim \frac{1}{N} \,.
\end{equation}

Both coherent and squeezed states are Gaussian ones, that is their wave functions are Gaussian in the position and momentum representations. It could be expected that the Heisenberg limit can be exceeded by non-Gaussian quantum states of light. However, this is not the case: the $1/N$ scaling of the phase sensitivity holds for arbitrary quantum states \cite{Ou_PRL_77_2352_1996, Giovannetti_PRL_108_210404_2012, Demkowicz_PIO_60-345_2015}.  

The common feature of Eqs.\,(\ref{SNL}-\ref{HL}) is that they describe the mean squared error of the measurement of an apriori unknown phase shift. At the same time, in many cases the measurement goal can be formulated in a different way --- namely, the binary (yes/no) detection of some given phase shift. In this case, the non-Gaussian quantum states could provide radically better detection fidelity than the Gaussian ones because they can be orthogonal to each other and therefore can be discriminated unambiguously \cite{HelstromBook}. 

Here we propose a scheme of interferometric measurement which implements this approach. In the next section, we describe this scheme and calculate its sensitivity in the ideal detection case. In Sec.\,\ref{sec:nonideal} we take into account of the detection inefficiency. In the concluding Sec.\,\ref{sec:conclusion}, the prospects of the experimental implementation of the non-Gaussian interferometer are discussed.

\section{Unambiguous phase shift detection scheme}

\begin{figure}
  \includegraphics[scale=1.2]{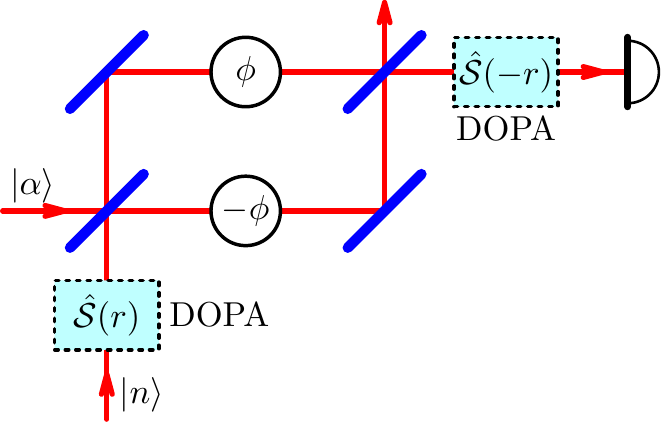}\quad
  \includegraphics[scale=1.2]{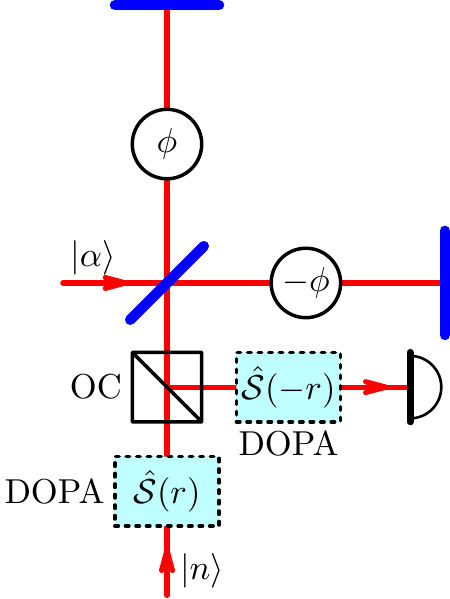}
  \caption{Two equivalent implementations of the non-Gaussian interferometer: the Mach-Zehnder (left) and the Michelson (right). OC --- optical circulator; DOPA --- optional degenerate optical parametric amplifiers.}\label{fig:scheme}
\end{figure}

Consider a standard two-arm (Mach-Zehnder or Michelson) interferometer, see Fig.\,\ref{fig:scheme}. Suppose that it is pumped through one of its input ports (the ``bright'' one) by classical (coherent) light. Suppose also that a single-photos optical pulse is injected into the second input port (the ``dark'' one). We assume that the interferometer is tuned in such a way that in the absence of a signal differential phase shift $\phi$ in its arms the classical pumping power is absent at one of its output ports, the ``dark'' one (the ``dark fringe'' regime). 

Let $\hat{{\rm a}}$, $\hat{{\rm b}}$ be the annihilation operators of the optical fields at, respectively, the bright and the dark input ports. We do not take into account for a while the optional degenerate optical parametric amplifiers shown in Fig.\,\ref{fig:scheme}, assuming that $r=0$. In this case, a straightforward calculation (using the Heisenberg picture) gives that the corresponding annihilation operator of the optical field at the output dark port is equal to 
\begin{equation}
  \hat{{\rm c}} = \hat{{\rm b}}\cos\phi + i(\alpha+\hat{{\rm a}})\sin\phi \,,
\end{equation} 
where $\alpha$ is mean coherent amplitude of input field, which we assume to be real (evidently, this assumption does not violate generality). 

For simplicity, we consider here the most important practically case of the strong classical pumping and the small signal phase shift, assuming that $\phi\to0$ and $\alpha\to\infty$, while the product $\alpha\phi$ remains finite. In this case, 
\begin{equation}\label{c_ideal} 
  \hat{{\rm c}} \to \hat{{\rm b}} + \beta 
  = \hat{\mathcal{D}}^\dagger(\beta)\hat{{\rm b}}\hat{\mathcal{D}}(\beta)\,,
\end{equation} 
where 
\begin{equation}
  \hat{\mathcal{D}}(\beta) = e^{\beta\hat{{\rm b}}^\dagger - \beta^*\hat{{\rm b}}}
\end{equation} 
is the displacement operator and 
\begin{equation}
  \beta = i\alpha\phi \,.
\end{equation} 

Suppose that an $n$-photon quantum state $\ket{n}$ in injected into the ``dark'' input port. In the Schroedinger picture, Eq.\,\eqref{c_ideal} corresponds to the following relation between this state and the corresponding state at the output dark port:
\begin{equation}\label{out_beta} 
  \ket{\beta,n} = \hat{\mathcal{D}}(\beta)\ket{n} \,.
\end{equation} 
In the absence of the signal, $\phi=0$, the input state just passes through the interferometer unchanged:
\begin{equation}\label{out_0} 
  \ket{0,n} = \ket{n} \,.
\end{equation} 
The scalar product of the states \eqref{out_0} and \eqref{out_beta} is equal to
\begin{equation}\label{scalar_prod} 
  \bracket{0,n}{\beta,n} = \bra{n}\mathcal{D}(\beta)\ket{n}
  = L_n(|\beta|^2)e^{-|\beta|^2/2} \,,
\end{equation} 
where $L_n$ is the $n$-th Laguerre polynomial. If $|\beta|^2$ is equal to one of its  roots:
\begin{equation}\label{orthog} 
  L_n(|\beta|^2) = 0 \,,
\end{equation} 
then the product \eqref{scalar_prod} is equal to zero, that is the states \eqref{out_0} and \eqref{out_beta} are orthogonal to each other. This means that these states can be discriminated unambiguously \cite{HelstromBook}.

As a practical measurement procedure, the straightforward counting of the photons number at the output dark port can be used. Really, it follows from Eqs.\,(\ref{out_beta}-\ref{scalar_prod}) that if the condition \eqref{orthog} is fulfilled then the output number of photons cannot be equal to $n$. At the same time, in the absence of the signal $(\beta=0)$, always exactly $n$ photons will be detected. 

The standard figures of merit of the binary detection schemes are the ``false negative'' (missing the signal) and ``false positive''  (erroneous detection) probabilities $P_{\rm f.n.}$ and $P_{\rm f.p.}$. In our case, $P_{\rm f.p.}$ evidently equal to zero, and 
\begin{equation}
  P_{\rm f.n.} = |\bracket{n}{\beta,n}|^2 = |L_n(|\beta|^2)|^2e^{-|\beta|^2} \,.
\end{equation}
In the practically most important case of the single-photon input state, $n=1$, the ``false negative'' probability is equal to 
\begin{equation}\label{P_fn_1} 
  P_{\rm f.n.} = (1-|\beta|^2)^2e^{-|\beta|^2} \,.
\end{equation}
and vanishes at $|\beta|=1$, that is at 
\begin{equation}\label{phi_opt} 
  |\phi| = \frac{1}{|\alpha|} \,. 
\end{equation} 

It is instructive to compare these results with the ones for the vacuum state at the input dark port. The corresponding output state is the coherent one:
\begin{equation}
  \ket{\beta,0} = \hat{\mathcal{D}}(\beta)\ket{0} \,,
\end{equation} 
which gives the Poissonian statistics of the detected number of quanta:
\begin{equation}
  |\bracket{n}{\beta,0}|^2 = \frac{|\beta|^{2n}e^{-|\beta|^2}}{n!} \,. 
\end{equation} 
In this case, $P_{\rm f.p.}=0$ and 
\begin{equation}\label{P_fn_0}
  P_{\rm f.n.} = e^{-|\beta|^2} \,.
\end{equation}
It is easy to see that the value of $|\beta|=1$ gives a significant ``false negative'' probability, equal to $1/e$.

\begin{figure}
  \includegraphics[width=0.6\textwidth]{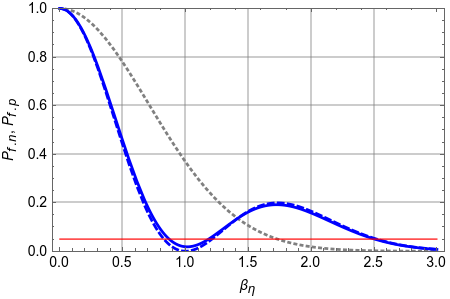}
  \caption{Plots of the error probabilities for the vacuum and the single-photon input states. Dots: $P_{\rm f.n.}$, vacuum input, $\eta=1$; dashes: $P_{\rm f.n.}$, single-photon input,  $\eta=1$; solid: $P_{\rm f.n.}$, single-photon input, $\eta=0.95$; thin horizontal line: $P_{\rm f.p.}$, single-photon input, $\eta=0.95$.}\label{fig:plots} 
\end{figure} 

In Fig.\,\ref{fig:plots}, the ``false negative'' probabilities \eqref{P_fn_1} and \eqref{P_fn_0} are plotted as functions of $|\beta|$. 

The optimal value of the signal phase shift \eqref{phi_opt} scales with $N$ as the shot noise limit \eqref{SNL}. However, similar to the Gaussian interferometry case, this scaling can be improved by using the additional squeezing of the incident quantum state. Suppose that before entering the interferometer, the $n$-photon pulse passes through the degenerate optical parametric amplifier (the squeezer), as shown in Fig.\,\ref{fig:scheme}. This gives the squeezed Fock state $\hat{\mathcal{S}}(r)\ket{n}$ at the interferometer input, where
\begin{equation}
  \hat{\mathcal{S}}(r) = e^{r(\hat{{\rm b}}^\dagger{^2} - \hat{{\rm b}}^2)/2}
\end{equation} 
is the squeeze operator. The corresponding quantum states at the interferometer output for the cases of $\phi\ne0$ and $\phi=0$ are equal to, respectively:
\begin{gather}
  \ket{\beta,r,n} = \hat{\mathcal{D}}(\beta)\hat{\mathcal{S}}(r)\ket{n} \label{out_beta_r}
    \,,\\
  \ket{0,r,n} = \hat{\mathcal{S}}(r)\ket{n} \label{out_0_r}  \,,
\end{gather}
compare with Eqs.\,(\ref{out_beta}, \ref{out_0}). Then note that 
\begin{equation}
  \hat{\mathcal{S}}^\dagger(r)\hat{\mathcal{D}}(\beta)\hat{\mathcal{S}}(r)
    = \hat{\mathcal{D}}(\beta e^r) \,.
\end{equation} 
Therefore, the scalar product of the states (\ref{out_0_r}, \ref{out_beta_r}) is equal to
\begin{equation}
  \bracket{0,r,n}{\beta,r,n} = \bra{n}\mathcal{D}(\beta e^r)\ket{n}
  = L_n(|\beta|^2e^{2r})e^{-|\beta|^2e^{2r}/2} \,,
\end{equation} 
Comparison of this equation with Eq.\,\eqref{scalar_prod} shows that the squeezing shrinks the values of $\beta$ which provide the unambiguous detection by the squeeze factor $e^{r}$. In particular, the condition \eqref{phi_opt} for the case of $n=1$ modifies as follows:
\begin{equation}\label{phi_opt_r} 
  |\phi| = \frac{e^{-r}}{|\alpha|} \,,
\end{equation} 
which corresponds to the squeezing-improved SNL \eqref{dphi_sqz}.

The quantum states (\ref{out_beta_r}, \ref{out_0_r}) can be discriminated using the following two-stage procedure. First, they have to be ``unsqueezed'' back using the second DOPA at the output of the interferometer, with the squeeze factor opposite to the one of the first one, see Fig.\,\ref{fig:scheme}:
\begin{gather}
  \hat{\mathcal{S}}(-r)\ket{\beta,r,n} 
    = \hat{\mathcal{S}}^\dagger(r)\hat{\mathcal{D}}(\beta)\hat{\mathcal{S}}(r)\ket{n}
    = \ket{\beta e^r,n} \,, \\
  \hat{\mathcal{S}}(-r)\ket{0,r,n} 
    = \hat{\mathcal{S}}^\dagger(r)\hat{\mathcal{S}}(r)\ket{n} = \ket{0,n} \,.
\end{gather}
Taking into account that these quantum states again has the form (\ref{out_beta}, \ref{out_0}), only with the amplified signal displacement $\beta e^{2r}$, the photons number counting still can be used to discriminate them.

Note that this layout, namely the interferometer preceded by the squeezer and followed by the anti-squeezer, was first proposed by C.Caves \cite{Caves1981} as the mean for suppressing of the photodetection inefficiency influence in the ``ordinary'' (Gaussian light) interferometers. As it is shown in the next section, the second DOPA provides the same advantage in the considered here non-Gaussian case as well.

\section{Photodetection inefficiency}\label{sec:nonideal}

It is well known that the sensitivity gain provided non-classical states of light is vulnerable to the optical losses. In the modern interferometers, the most serious limiting factor is the photodetection inefficiency, see \eg the loss budget analysis in \cite{Nature_2011, Nature_2013}). Here we calculate how this inefficiency affects the non-Gaussian interferometry. 
\begin{figure}
  \includegraphics[scale=1.2]{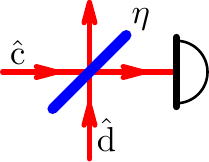}
  \caption{Model of a non-ideal detector with the quantum efficiency $\eta$}\label{fig:lossy_det}
\end{figure}

We model the finite quantum efficiency $\eta<1$ of the detector by means of an imaginary beamsplitter with the power transmissivity $\eta$, which mixes the measured field $\hat{{\rm b}}$ with the auxiliary vacuum field described by the annihilation operator $\hat{{\rm d}}$, see Fig.\,\ref{fig:lossy_det}. In the Heisenberg picture, this beamsplitter transforms its incident fields as follows:
\begin{subequations}
  \begin{gather}
    \hat{\mathcal{U}}^\dagger\hat{{\rm c}}\hat{\mathcal{U}} 
      = \sqrt{\eta}\,\hat{{\rm c}} + \sqrt{1-\eta}\,\hat{{\rm d}} \,, \\
    \hat{\mathcal{U}}^\dagger\hat{{\rm d}}\hat{\mathcal{U}} 
      = -\sqrt{1-\eta}\,\hat{{\rm c}} + \sqrt{\eta}\,\hat{{\rm d}} \,,
  \end{gather}
\end{subequations}
where $\hat{\mathcal{U}}$ is the corresponding unitary evolution operator. It follows from these equations, that
\begin{equation}
    \hat{\mathcal{U}}\hat{{\rm c}}\hat{\mathcal{U}}^\dagger 
      = \sqrt{\eta}\,\hat{{\rm c}} - \sqrt{1-\eta}\,\hat{{\rm d}} \,, \\
\end{equation} 

In the Schroedinger picture, the transformation of the quantum state $\hat{\rho}$ at the detector input to the effective damped one $\hat{\rho}_\eta$ is described by the following equation:
\begin{equation}
  \hat{\rho}_\eta = \Tr_d\bigl(
      \hat{\mathcal{U}}\hat{\rho}\otimes\ket{0}_d\,{}_d\bra{0}\hat{\mathcal{U}}^\dagger
    \bigr) ,
\end{equation} 
where $\ket{0}_d$ is ground state wave function of the auxiliary optical mode and the trace is taken over the states of this mode. 

In order to avoid over-cluttering of the equations, we limit ourselves by the simplest but the most important case of the single-quantum state:
\begin{equation}
  \hat{\rho} = \ket{\beta e^r,1}\bra{\beta e^{r}, 1} \,.
\end{equation} 
It can be shown that in this case,
\begin{equation}\label{rho_beta} 
  \hat{\rho}_\eta = \eta\ket{\beta_\eta,1}\bra{\beta_\eta, 1}
    + (1-\eta)\ket{\beta_\eta,0}\bra{\beta_\eta,0} \,,
\end{equation} 
where
\begin{equation}
  \beta_\eta = \sqrt{\eta}\beta e^r \,.
\end{equation} 

It follows from Eq.\,\eqref{rho_beta} that the detection inefficiency affects the sensitivity in several ways. First, it reduces the effective signal by the factor $\sqrt{\eta}$. However, for the state-of-art detectors with $1-\eta\ll1$ the corresponding decrease of $\beta_\eta$ is small and, if necessary, can be compensated by the proportional increase of the pump power or squeeze factor. Second, the detection inefficiency dilutes the non-Gaussian state $\ket{\beta_\eta, 1}$ with the coherent (Gaussian) state $\ket{\beta_\eta,0}$. The resulting ``false negative'' probability for the state \eqref{rho_beta} is equal to
\begin{equation}\label{P_fn_eta} 
  P_{\rm f.n.} = \bra{1}\hat{\rho}_\eta\ket{1}
  = \bigl[\eta\bigl(1 - |\beta_\eta|^2\bigr)^2 + (1-\eta)|\beta_\eta|^2\bigr]
      e^{-|\beta_\eta|^2} \,.
\end{equation} 
The minimum of this equation in $\beta_\eta$ is provided by 
\begin{equation}\label{opt_beta_eta} 
  |\beta_\eta|=1 
\end{equation} 
and is equal to
\begin{equation}\label{P_fn_eta}
  P_{\rm f.n.} = \frac{1-\eta}{e} \,.
\end{equation} 
Therefore, the unambiguous detection is impossible in this case. However, the ``false negative'' probability is suppressed by $1-\eta$ in comparison with the Gaussian case, compare Eqs.\,\eqref{P_fn_0} and \eqref{P_fn_eta}. The detection inefficiency creates also a non-zero ``false positive'' probability:
\begin{equation}\label{P_fp_eta} 
  P_{\rm f.p.} = 1 - \bra{1}\hat{\rho}_\eta\ket{1}\bigr|_{\rm \beta=0} = 1-\eta \,.
\end{equation} 
The error probabilities (\ref{P_fn_eta}, \ref{P_fp_eta}) are plotted in Fig.\,\ref{fig:plots} as functions of $\beta_\eta$ for the demanding but realistic case of $\eta=0.95$ \cite{Lita_OE_16_3032_2008}. 

Note that these error probabilities do not depend on the squeeze factor $r$ (for the given value of $\beta_\eta$), that is the sensitivity degradation imposed by the detection inefficiency is not sensitive to the squeezing degree of the input quantum state. The reason for this is the amplification, provided by the second DOPA \cite{Caves1981}. 

\section{Conclusion}\label{sec:conclusion} 

The scheme of the non-Gaussian interferometer considered here crucially depends on two additional (in comparison with ordinary interferometers) elements: (i) a deterministic source of source of single-photon light pulses and (ii) a high-efficient photon-number-resolving photodetector. Fortunately, technologies of both generation and detection of single photons were under active development during the last decades, stimulated by the needs of quantum information science, see \eg the review articles \cite{Grangier_NJP_6_1_2004, Lounis_RPP_68_1129_2005, Eisaman_RSI_82_071101_2011}, as well as the recent work \cite{Chu_NPhoton_11_58_2016}. 

In particular, the photon-number-resolving quantum efficiency exceeding 95\% was demonstrated using the modern cryogenic transition-edge sensors \cite{Lita_OE_16_3032_2008}. It can be seen from Eqs.\,(\ref{P_fn_eta}, \ref{P_fp_eta}) and Fig.\,\ref{fig:plots} that for this value of the quantum efficiency, the gain in the detection error probability approaches one order of magnitude, in comparison with the ordinary Gaussian-state interferometry. Moreover, the error probability is noticeably suppressed not only at the optimal value of the phase shift, defined by Eq.\,\eqref{opt_beta_eta}, but also in the rather broad range around this value. This means that the considered scheme is robust with respect to deviations of the signal phase shift from the apriory value \eqref{opt_beta_eta}.  

\acknowledgments

The author acknowledges funding by the RFBR grant 16-52-12031.


\end{document}